%% file: main.tex
\documentclass{sig-alternate-10pt}
\usepackage[bookmarks=false]{hyperref}
\usepackage[yyyymmdd,hhmmss]{datetime}
\usepackage{verbatim}
\usepackage{subcaption}
\usepackage{caption}

\usepackage{cite}

\usepackage[printonlyused,withpage]{acronym}
\usepackage{xcolor}
\usepackage{amssymb}
\usepackage{graphicx}
\usepackage{fixme}
\fxsetup{
status=draft,
    layout=inline,
    theme=color,
    mode=multiuser
}
\FXRegisterAuthor{pn}{pnn}{\color{red}PN}

\setkeys{Gin}{draft=false}

\begin{document}

\title{Performance Evaluation of MU-MIMO Under the Impact of Open Loop Traffic Dynamics}

\author{Peshal Nayak \\
ECE Department, \\ 
Rice University, \\
Houston, TX, USA
}
\maketitle


\input{secs/abstract}%
\input{secs/introduction}%
\input{secs/openloop}
\input{secs/background}

\input{secs/conclusion}
\bibliographystyle{unsrt}
\bibliography{main}

\end{document}

%% file: secs/abstract.tex
\begin{abstract}
\boldmath Multi-user MIMO enhances throughput by simultaneously transmitting/receiving parallel data streams to/from a group of users. However, the throughput analysis does not account for variable data traffic. The research objective of this project is to analyze the effect of variable traffic as it will change the system behaviour. The two characteristic features of variable traffic viz. packet size variation and traffic burstiness have been considered for the analysis in this project. Their effects have been studied individually to provide an insight into the problem. Through simulations, we show that in certain scenarios the system performance of MU-MIMO deteriorates significantly due to both packet size variations and traffic burstiness individually even under ideal conditions with respect to channel variations, channel correlation, mobility, etc. Furthermore, we show that under bursty traffic with higher offered load the aggregate throughput first remains steady until a certain peak to average rate ratio and then deteriorates linearly instead of exponentially. Also, under high amount of traffic burstiness, the throughputs are independent of the aggregation rates. A thorough analysis is provided to explain these two phenomena. This is the first research work that considers the impact of variable traffic on the performance of MU-MIMO. Therefore, the implications for MAC protocol design are significant.
\end{abstract}


%% file: secs/introduction.tex
\section{Introduction}
\label{sec:intro}

Multi-user MIMO improves system performance by transmitting independent data streams to multiple users in parallel. For an access point with multiple antennas, the transmission process involves selection of users based on a specific criteria such that there is low spatial correlation between them followed by precoding (e.g beamforming) the data streams to achieve multi-user interference cancellation. However, such precoding schemes need the channel state information (CSI) for different users which the access point has to acquire prior to data transmission through the process of channel sounding. During channel sounding, a training sequence is sent to intended users on the downlink which is used by the receivers to estimate the CSI. The receivers then transmit the CSI back to the AP during the channel feedback stage. This CSI is then used to compute beam-steering weights required to perform downlink beamforming.  

Although this scheme may improve the throughput, there is a considerable overhead involved for setting up communication. Though the overhead can be amortised by increasing the size of the data transmitted, packet size variations might make it difficult to gather such large sized data. Furthermore, no performance analysis has accounted for bursty traffic. For instance, all the users in the network may not be infinitely backlogged. In such a scenario, it may be impossible to group users together and transmitting multiple streams parallely may not be possible at all. 

The ultimate target of this project is to analyse the effect of traffic on performance of Multi-user WLANs and resulting implications for MAC layer decisions. This report presents an analysis of the performance of MU-MIMO under variable traffic. Through simulations and thorough analysis we show that the significant performance gains promised by MU-MIMO are severely mitigated by variable traffic. We show that a general MU-MIMO MAC protocol which is not sensitive to variable traffic causes performance degradation even under ideal channel conditions. Prior works related to performance evaluation or enhancement have never considered such effects. This report is organized as follows. In Section 2, we evaluate each performance factor by first stating the hypothesis, followed a description of the simulation setup, the simulation results and finally an analysis of the results. 

%% file: secs/openloop.tex
\section{Effect of Variable Traffic}
\begin{figure}[tbh]
  \centering
    \includegraphics[width=0.5\textwidth]{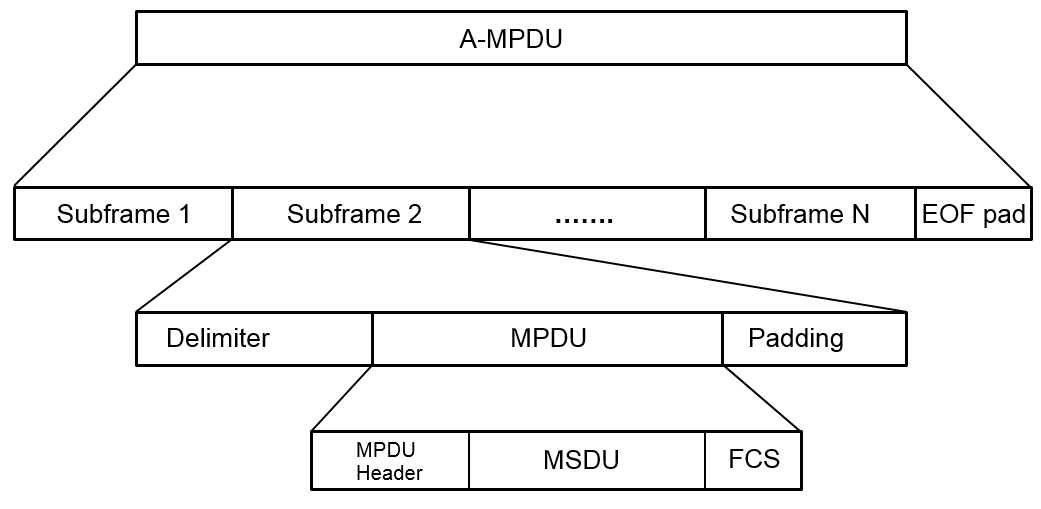}
    \caption{MAC Protocol Data Unit Aggregation as stated in 802.11ac [2]}
    \label{fig1}
\end{figure}

Two of the important characteristics of traffic considered in this analysis are packet size variation and burstiness. For setting up communication between AP and STAs in case of MU-MIMO, a large amount overhead has to be incurred. It is possible to bear the cost of the overhead by transmitting large amounts of data, thereby gaining considerable amount of gains in terms of throughput due to multi-user transmission. In order to transmit greater amounts of data per transmit cycle, frame aggregation is used which reduces the overhead to data ratio. For the purpose of analysing the effect of aggregation on the system performance, the specifications of the 802.11 ac \cite{2} standard have been used. The frame aggregation technique used in 802.11 ac is MAC Protocol Data Unit Aggregation (A-MPDU). As shown in fig.~\ref{fig1} sub-MPDU frames are formed by appending a MAC header and FCS to them. Multiple such subframes are concatenated by inserting a MPDU delimiter at the start and padding bits at the end of each subframe to form a Physical Service Data Unit (PSDU) which is transmitted by the AP. It is worth stating here that 802.11 ac supports aggregation rates upto 64.

\subsection{Packet Size Variation}
\subsubsection{Issue with variable packet size}
For MU-MIMO as data is transmitted parallely to different users, the transmit times for one user would affect the other users grouped with it and vice versa. There could be two cases: the packet sizes for all users could be the same resulting in same length data streams as shown in fig~\ref{fig2.1}. However, due to packet size variation, the transmit times for different users might be different as shown in fig.~\ref{fig2.2}. As a result, users with smaller transmit times would suffer. If a fragmentation scheme is used in this situation to adjust the transmit times, the result would be a higher overhead to data ratio which would further hamper throughput.

The size of packets transmitted over the internet typically follows bimodal distribution with one peak at maximum packet size and the other close to minimum packet size. In \cite{1}, John and Tafvelin studied the packet size distributions for IPv4 packets and reported that the bimodal distribution had around 44\% of the packets lengths less than 100 bytes while another 37\% were between 1400 and 1500 (the MTU in case of Ethernet). Similar results were also reported by \cite{nayak2021ap, nayak2021uscope, nayak2019virtual, nayakpassive}.

\begin{figure}
        \centering
        \begin{subfigure}[b]{0.5\textwidth}
                \includegraphics[width=\textwidth]{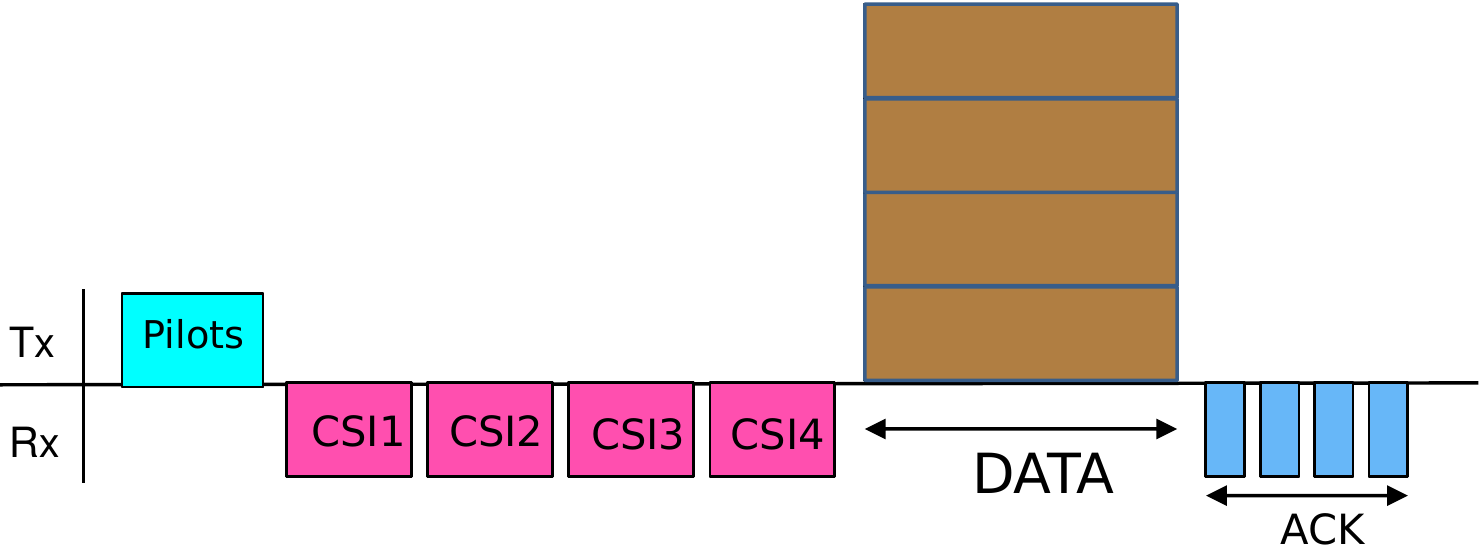}
                \caption{Same length data streams due to fixed packet size}
                \label{fig2.1}
        \end{subfigure}%
        \\
        ~ 
        \begin{subfigure}[b]{0.5\textwidth}
                \includegraphics[width=\textwidth]{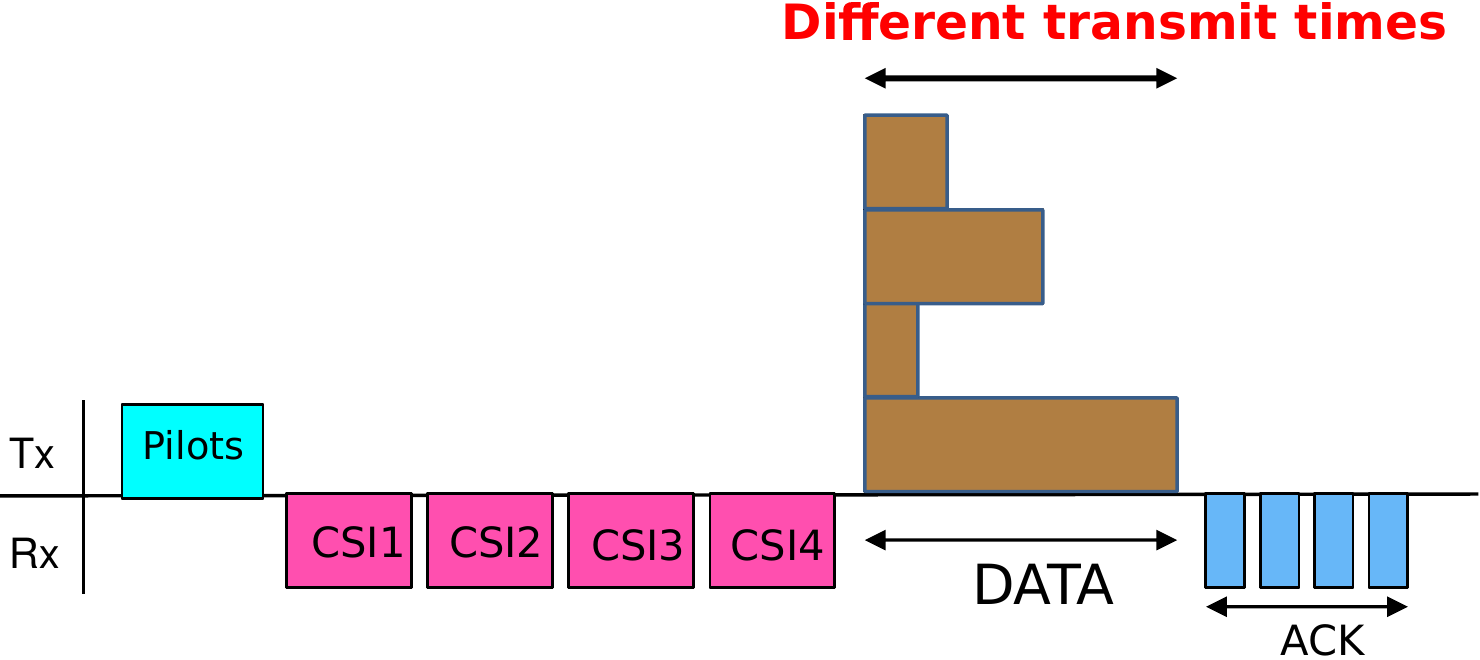}
                \caption{Variable length data streams due to packet size variation. For users with lower transmit time, the remaining part of data transmission phase is wasted.}
                \label{fig2.2}
        \end{subfigure}
        \caption{Examples of timing diagram for MU-MIMO under variable traffic}\label{fig2}
\end{figure}  
Variation of packet size does affect of system performance in case of SISO systems as well. However, as data is not sent simultaneously, a particular user's transmit time does not affect others. Users with longer transmit times do end up using most of the network resources. But that problem exists in case of MU-MIMO as well and can be solved by a throughput fairness scheme. Also, the overhead involved in SISO is significantly less as compared to MU-MIMO and can be amortized even at lower aggregation rates. Collisions and channel variation causes resource wastage if the number of aggregated frames is large. The problem due to channel variation would be significant in case of MU-MIMO specially because it would lay restrictions on the validity period of the CSI obtained from the stations during channel state feedback stage. As a result, larger aggregation rates might not be a good choice for MU-MIMO either. Besides, as shown later, the effect of packet size variation would become worse with increasing aggregation rates. 

\subsubsection{Simulation Setup}
This section describes the simulation setup and associated parameters and the different scenarios simulated. For an M antenna AP serving N users, every downlink transmission would require selection of `M' users based on some criteria followed by parallel transmission of data to these M users. The approach used in this analysis is to identify a performance factor and then to isolate its effect to the maximum possible extent. For studying the effect that  traffic dynamics has on multi-user MIMO performance, it is necessary to assume that other aspects of the system such as precoding, transmission and reception happen perfectly,  that no errors are introduced due to channel variation and that the receivers are stationary. Such assumptions help isolate and analyse the effect of traffic dynamics on the system performance. As the effect of packet size variation and traffic burstiness (described in the previous subsection) are independent of each other, they are studied separately. Thus, while analysing the effect of packet size variation, the AP is kept in a backlogged state with respect to all users and while analysing the effect of traffic burstiness, the packet sizes are kept fixed at the mean value.

The effect of packet size variation remains the same regardless of the number of transmit antennas and the number of users in the system so long as multiuser transmission is involved. Therefore, the case in which the number of users in the system equals the number of transmit antennas (M=N=4) is considered as it simplifies the simulation without affecting the final analysis. 

With respect to packet size variation, the best case scenario occurs when every user receives packets of fixed size (equal to the mean value) as shown in fig.~\ref{fig2.1}. However, as mentioned in the previous subsection, the packet sizes follow a bimodal distribution. By adding more weights on the two extreme modes, the packet size variation would approach a Bernoulli distribution in the worst case as shown in fig.~\ref{fig2.2}. However, by keeping the two extreme modes and the mean packet size fixed, a number of packet size distribution can be generated with variances between those of the two extreme cases. For the sake of representation, we use normalized variance which would be calculated as follows. The axis of packet size variation plot is normalized with respect to the maximum. Thus, instead of varying between 0 and 1024, the x-axis would now vary between 0 and 1. This makes it easy to represent variance as it now varies from 0 (best case scenario) to 0.25 (worst case scenario) making it easy to represent it on an axis. Note that this does not change the packet sizes in the simulations and is only for representational purposes.

\begin{figure}[tbh]
        \centering
        \begin{subfigure}[b]{0.46\textwidth}
                \includegraphics[width=\textwidth]{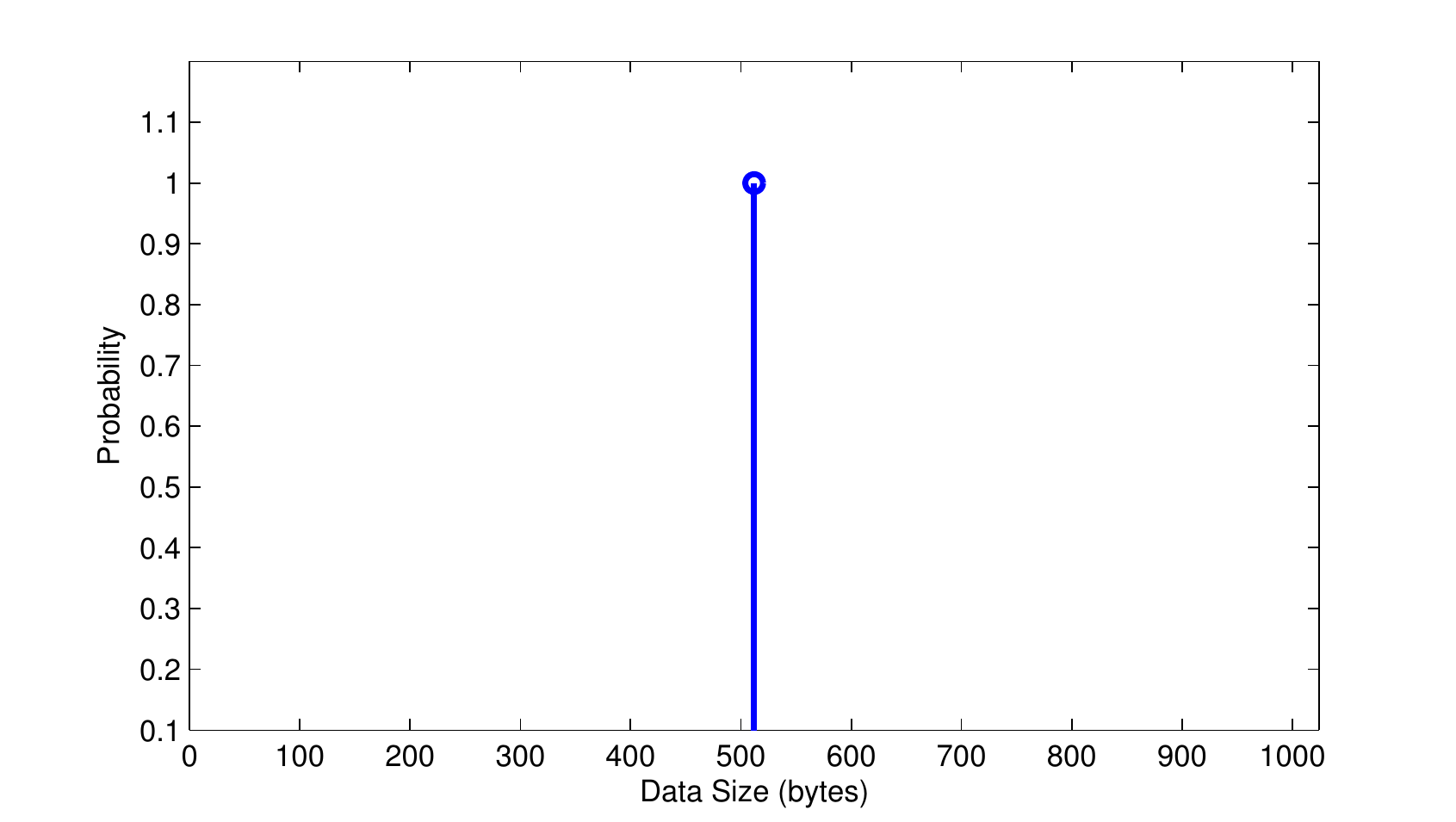}
                \caption{Best case scenario: Fixed mean sized packets}
                \label{fig2.1}
        \end{subfigure}%
        \\
        ~ 
        \begin{subfigure}[b]{0.46\textwidth}
                \includegraphics[width=\textwidth]{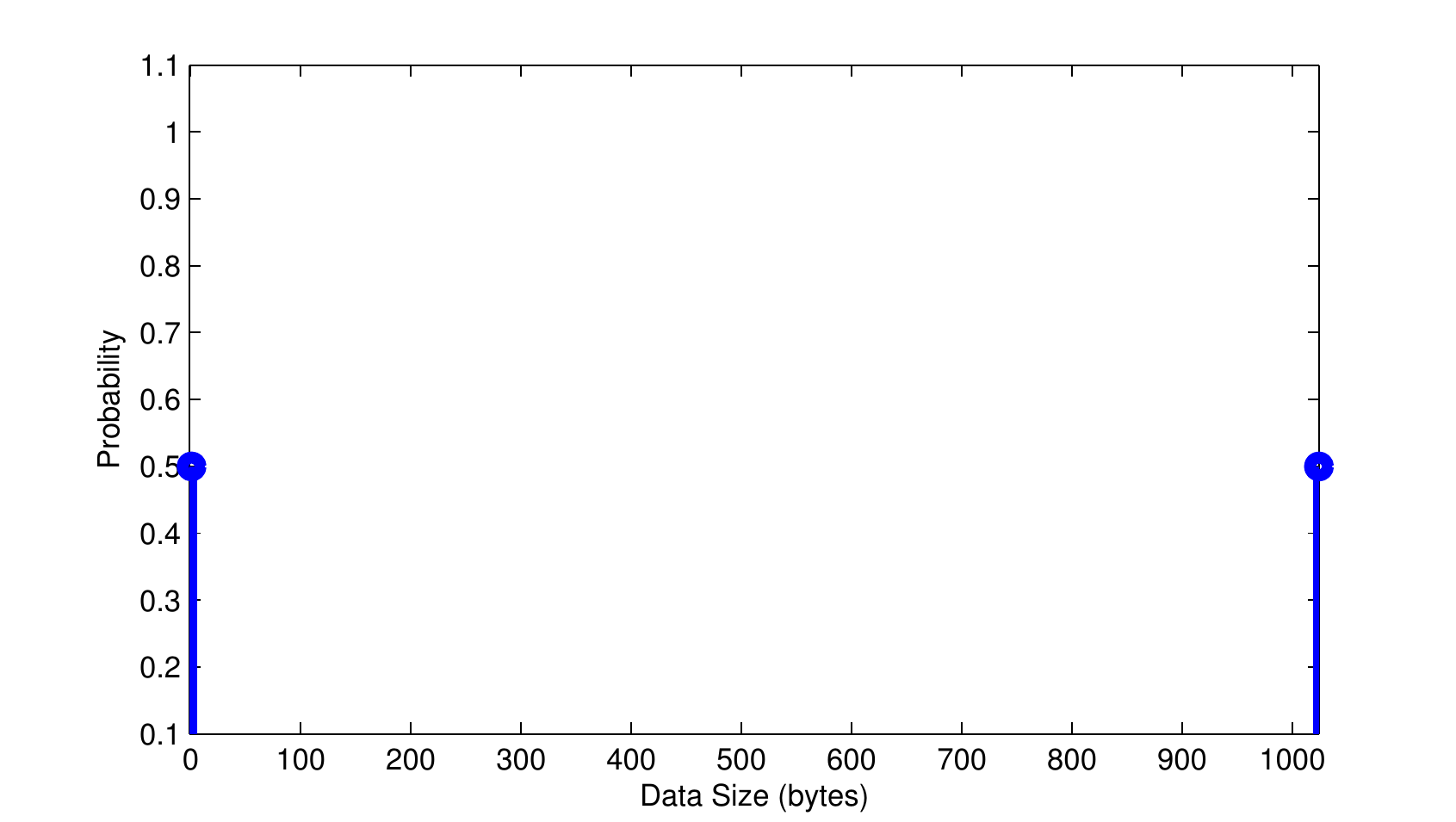}
                \caption{Worst case scenario: Bernoulli Distribution. Probability of the two modes is maximum}
                \label{fig2.2}
        \end{subfigure}
        \caption{Extreme cases for packet size variation}\label{fig2}
\end{figure}

\begin{figure}
        \centering
        \begin{subfigure}[b]{0.5\textwidth}
                \includegraphics[width=\textwidth]{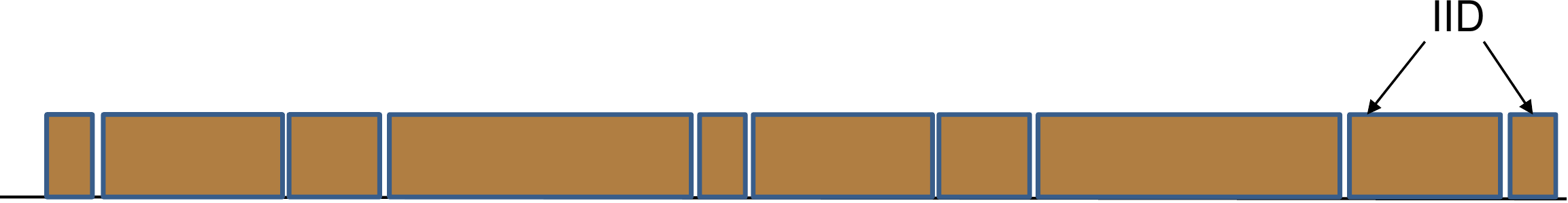}
                \caption{Best case file model: IID (Packet are generated from the same distribution but consecutive packet sizes are independent)}
                \label{fig3.1}
        \end{subfigure}%
        \hfill \\
        \hfill \\
        ~ 
        \begin{subfigure}[b]{0.5\textwidth}
                \includegraphics[width=\textwidth]{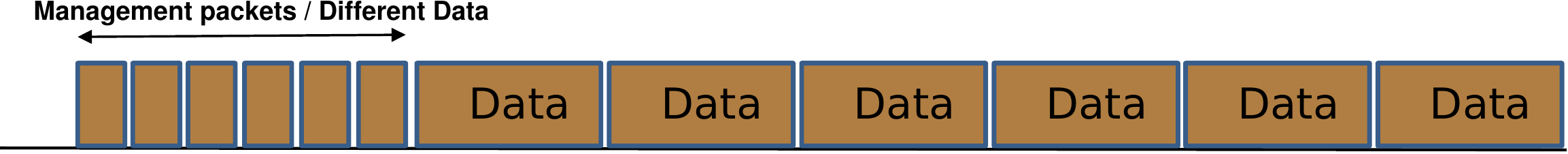}
                \caption{More realistic file model: Correlated (Packets sizes are similar to preceding packets) with correlation coefficient of 6}
                \label{fig3.2}
        \end{subfigure}
        \caption{Two different file models considered to generate packets}\label{fig3}
\end{figure}

For studying the effect of packet size variation, two types of file models are used:

\begin{itemize}
\item {\bf Independently and Identically Distributed Packet Sizes (IID):} In this model, the packet sizes are generated in an IID fashion i.e. each packet's size is independent of the size of the previous packets and all packets follow the same bimodal distribution. A simple illustration is shown in fig~\ref{fig3.1} This is a simple and best case file model.
\item {\bf Correlated Packet Sizes:} For real applications, packets are generally correlated. For instance, for file downloads most applications would generally send a certain number of data packets of larger sizes followed by certain application based management packets of smaller sizes or applications that have two types of data, for instance: Skype (video and audio). A simple illustration of correlated packet sizes is as shown in fig~\ref{fig3.2}. For correlated packet sizes, the correlation coefficient is defined as the number of correlated packets generated by the traffic source. 
\end{itemize}

For modelling the traffic source, an ON/OFF traffic source from ns-3 is used to generate traffic. As the effect of traffic burstiness is studied separately, the traffic source in this case is kept ON all the time. This causes the AP to be backlogged with packets for all users. In each of the above scenario, A-MPDU is used as the aggregation scheme. The maximum number of aggregated frames is kept at 40 frames. Though 802.11 ac allows higher number of aggregated frames, such higher number of aggregated frames would reduce the quality of service in case of certain applications such as voice. The parameters used for this simulation have been stated in TABLE~\ref{table1}. For the calculation of overhead, the bandwidth, the sub-carrier grouping and the quantization bits have been chosen to result in a minimum amount of overhead from channel sounding. The sizes for IP and UDP headers are the default sizes from ns-3. The values for other headers are as stated in 802.11 ac \cite{2}.

\begin{table}[tbh]
\caption{Parameters set for all Simulations: Packet size variation and traffic burstiness}
\label{table1}
\centering
\begin{tabular}{ |l|l|}
    \hline
    Bandwidth & 20 MHz \\ \hline
    Subcarrier Grouping & 4 \\ \hline
    $\psi$ , $\phi$ (Quantization bits) & 5,7 \\ \hline
    PHY header & 44 $\mu$s \\ \hline
    MAC header & 36 bytes \\ \hline
    Delimiter & 4 bytes \\ \hline
    FCS & 4 bytes \\ \hline
    Padding & 0-3 bytes \\ \hline
    EOF padding & 0-3 bytes \\ \hline
    MCS scheme & 54 Mbps \\ \hline    
    IP header & 20 bytes \\ \hline
    UDP header & 8 bytes \\ \hline
    \end{tabular}
\end{table}
 
\subsubsection{Simulation, Results and Analysis}
\begin{figure}[tbh]
  \centering
    \includegraphics[width=0.55\textwidth]{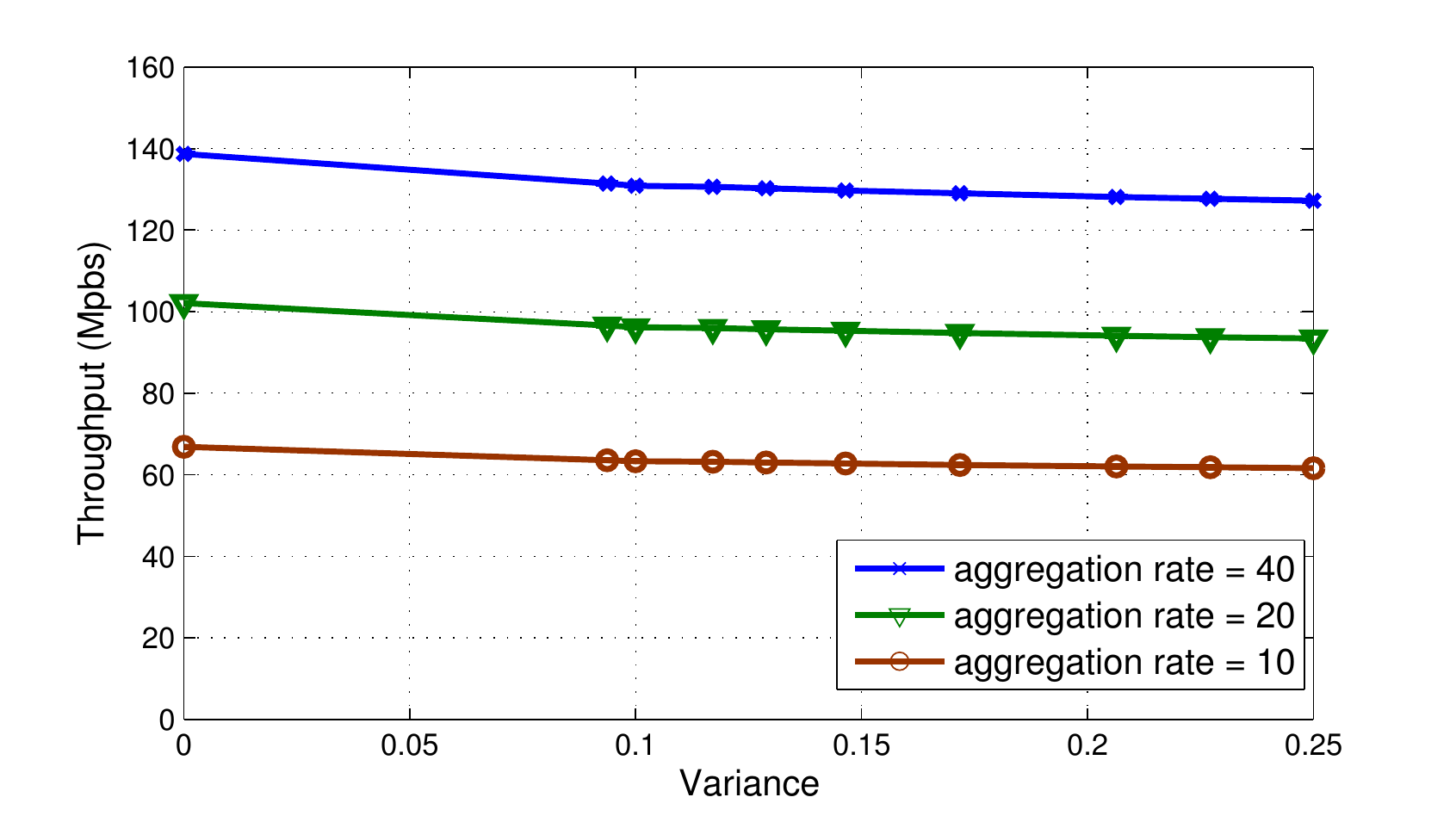}
    \caption{Effect of variance on the performance of the system. Zero variance denotes the best case scenario and 0.25 variance denotes the worst case}
    \label{fig5}
\end{figure}

\begin{figure}[tbh]
  \centering
    \includegraphics[width=0.55\textwidth]{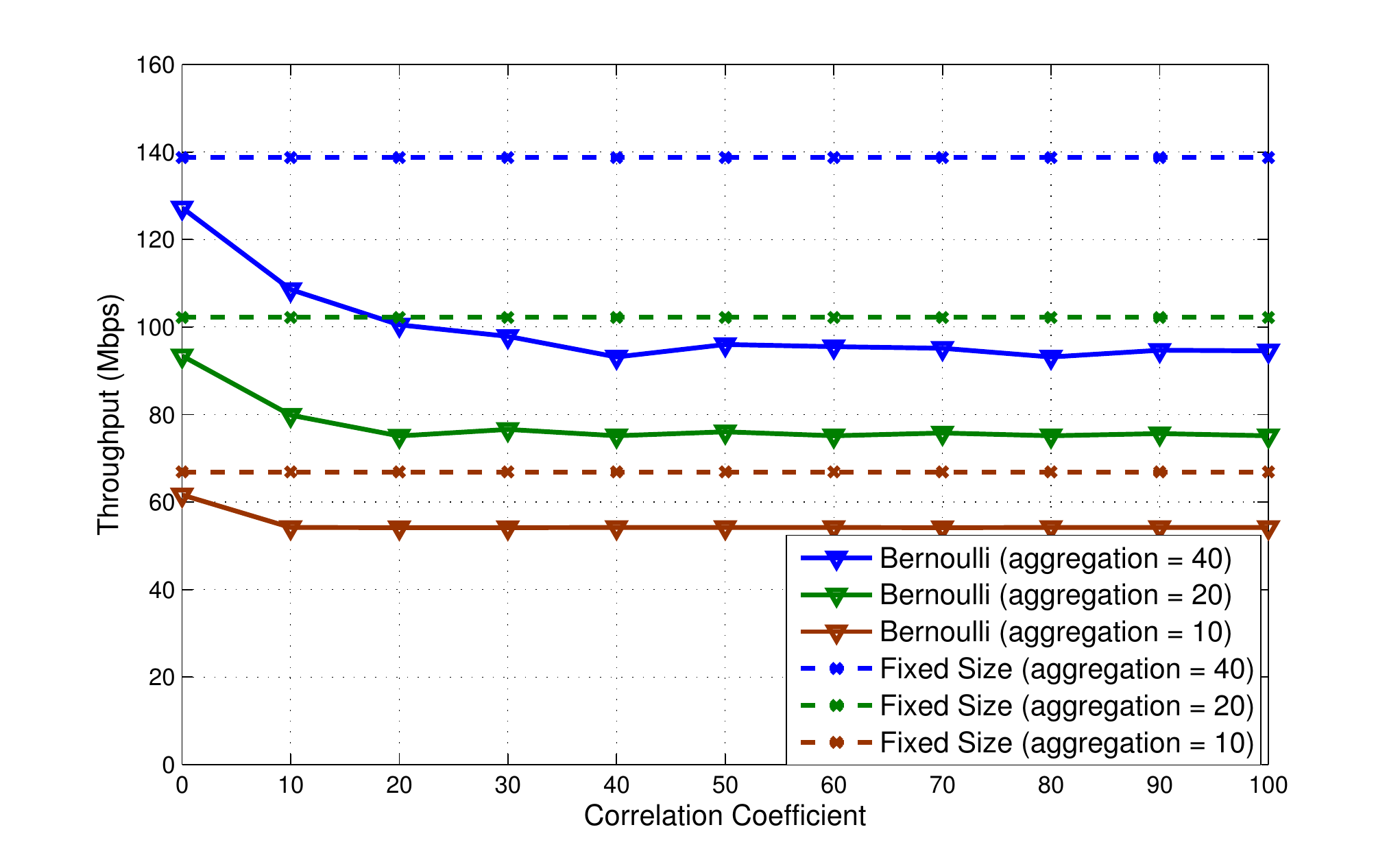}
    \caption{Effect of packet size variation on throughput. Zero correlation denotes IID case}
    \label{fig6}
\end{figure}

For packet size variation, in the case when the packets are generated in an IID fashion, the effect of changes in variance on throughput is as shown in fig. ~\ref{fig5}. As the variance changes from zero (best case scenario) to 0.25 (worst case scenario), the drop in throughput is very minimal (approximately 9 Mbps at the aggregation rate of 40). Also, this drop in throughput increases with increasing aggregation rate. However, as aggregation rate decreases the overhead to data ratio increases thereby decreasing the maximum achievable throughput (in the best case scenario). However, as seen in fig.~\ref{fig6}, for the case when packet sizes are correlated the decrement in throughput is very large. Again the drop in throughput becomes large at higher aggregation rate. This is because at lower aggregation rates, the variation in transmit times for different users is less as compared to the overhead in each transmit cycle. This difference becomes larger at higher aggregation rates causing a larger drop in throughput. Also, the curves saturate at higher correlation factors. The critical point for saturation is the point where the correlation coefficient becomes equal to the aggregation rate. This is because beyond this point the possible combinations of aggregated packets is the same even as the correlation coefficient is increased. Hence, the throughput remains more or less the same.

%
%
%

\subsection{Traffic Burstiness}
\subsubsection{Issue with bursty traffic}
For MU-MIMO, the AP generally groups users such that their channels are uncorrelated so that the rows of the channel matrix are linearly independent. However, measurement results have indicated that for WLANs the channels for users are generally uncorrelated in time and so the users can be grouped on the basis of other criteria to maximize their throughput. If the AP transmits packets on a first in first out (FIFO) basis, users would be grouped one the basis of availability of packets in the transmit queue. However, traffic in real networks is bursty and hence users might not be backlogged all the time. Traffic burstiness is characterized by peak to average rate ratio. If the peak to average rate ratio is high, packets would arrive in the form of a tiny burst followed by a long idle time. 

\begin{figure}[tbh]
        \centering
        \begin{subfigure}[b]{0.5\textwidth}
                \includegraphics[width=\textwidth]{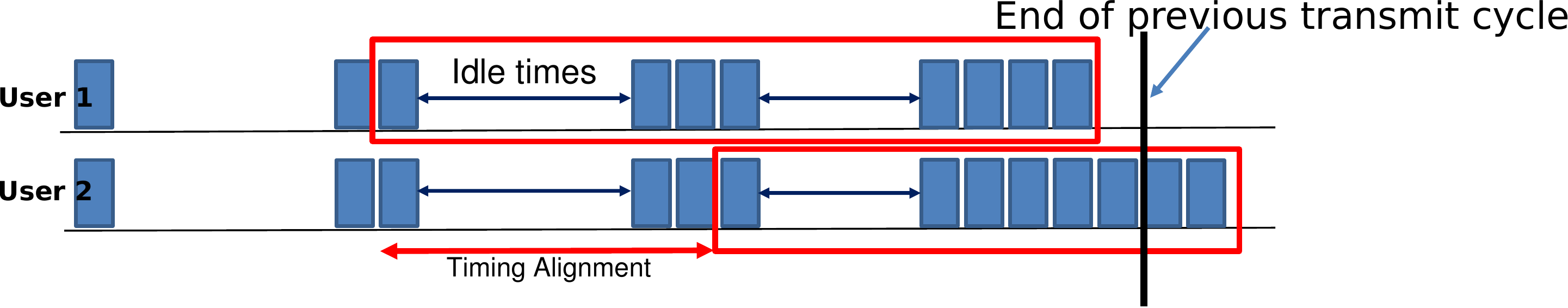}
                \caption{After previous transmit cycle the AP has to wait until it receives sufficient packets for both users}
                \label{fig4.1}
        \end{subfigure}%
        \hfill \\
        \hfill \\
        ~ 
        \begin{subfigure}[b]{0.5\textwidth}
                \includegraphics[width=\textwidth]{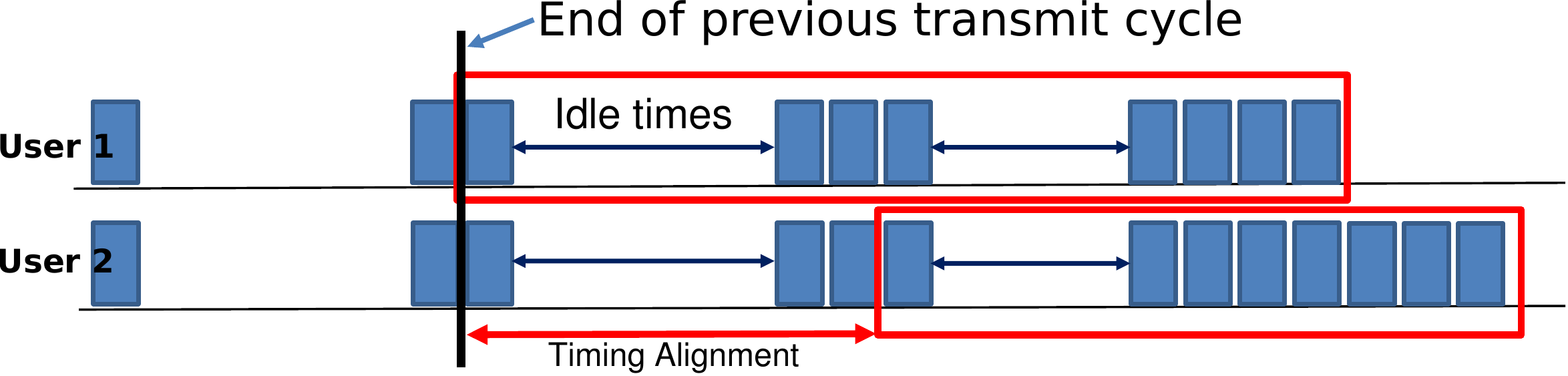}
                \caption{As the packets are already ready for transmission, the AP does not have to wait after the end of previous transmit cycle}
                \label{fig4.2}
        \end{subfigure}
        \caption{Examples for explaining the effect of bursty traffic on MU-MIMO system performance}\label{fig4}
\end{figure}

Bursty traffic affects the performance of single antenna systems as well. However, the throughput reduction is caused primarily due to a lower offered load caused by a high peak to average rate ratio. But there is no solution to this problem of lower offered load as the burstiness is mainly caused by applications running at the user end which cannot be influenced by the AP. However, packet arrivals in the form of bursts causes an additional problem of timing alignment even at higher offered loads. If an access point with M antennas does not have packets for M users, the delay in packet arrival could deteriorate the performance. The issue of timing alignment is shown for two user case in fig.~\ref{fig4} for an aggregation rate of 8. There are two cases described:
\begin{itemize}
\item {\bf Case1:} The AP has to wait until it receives sufficient packets for both users. This non-alignment might happen even a higher offered loads. And in this case the timing alignment issue will certainly affect the performance of the system due to delay in transmission. This is shown in fig.~\ref{fig4.1}.
\item {\bf Case2:} The AP does not have to wait for packets as they are already available. In this case, despite non-alignment, there is no effect on the system performance as there is no transmit delay. This is shown in fig.~\ref{fig4.2}.
\end{itemize}
For SISO, systems this problem is not encountered as the AP serves one user at a time and hence the throughput remains unaffected as long as the offered load is high enough.

\subsubsection{Simulation Setup}
Most of the general parameters of the simulator remain the same as before. This section describes only changes to the simulator for analysis of traffic burstiness. As the traffic burstiness increases the offered load decreases which causes the throughput to decrease. However, as mentioned in the previous sub-section, even at higher offered loads multi-user transmission faces a problem due to alignment of packets in time. To study the effect of non-alignment in time, the offered load into the system has to be maintained above the critical offered load. One way of doing this even at higher peak to average rate ratios is to increase the number of users. As the peak to average rate ratios is the same for all users (i.e. per user offered load is the same), as the number of users increase the offered load increases in the same proportion. The number of users in the simulation has been fixed at 12 which ensures that the offered load is high even at the maximum value of peak to average rate ratio considered in the simulation. In simulations studying the effect of traffic burstiness on the system performance, the packet sizes have been fixed at the mean value. Also, for modelling bursty traffic, the ON/OFF traffic source has been used with ON and OFF times following an exponential distribution. Thus, the mean ON and mean OFF times are used to control the peak to average rate ratio.

\subsubsection{Simulation Results}
\begin{figure*}[tbh]
        \centering
        \begin{subfigure}[b]{0.5\textwidth}
                \includegraphics[width=\textwidth]{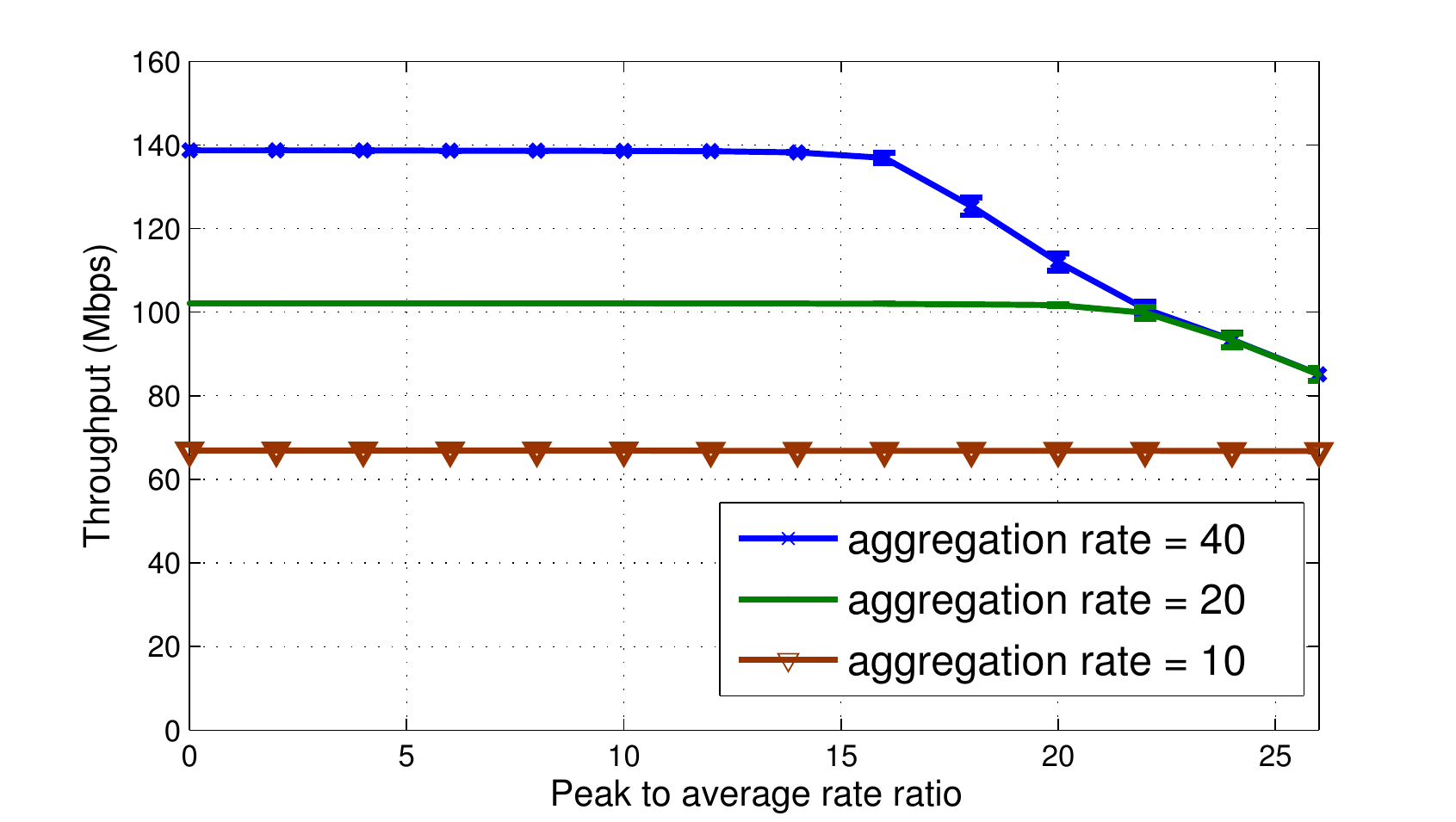}
                \caption{System performance deteriorates with increasing traffic burstiness}
                \label{fig7.1}
        \end{subfigure}%
        ~ 
        \begin{subfigure}[b]{0.5\textwidth}
                \includegraphics[width=\textwidth]{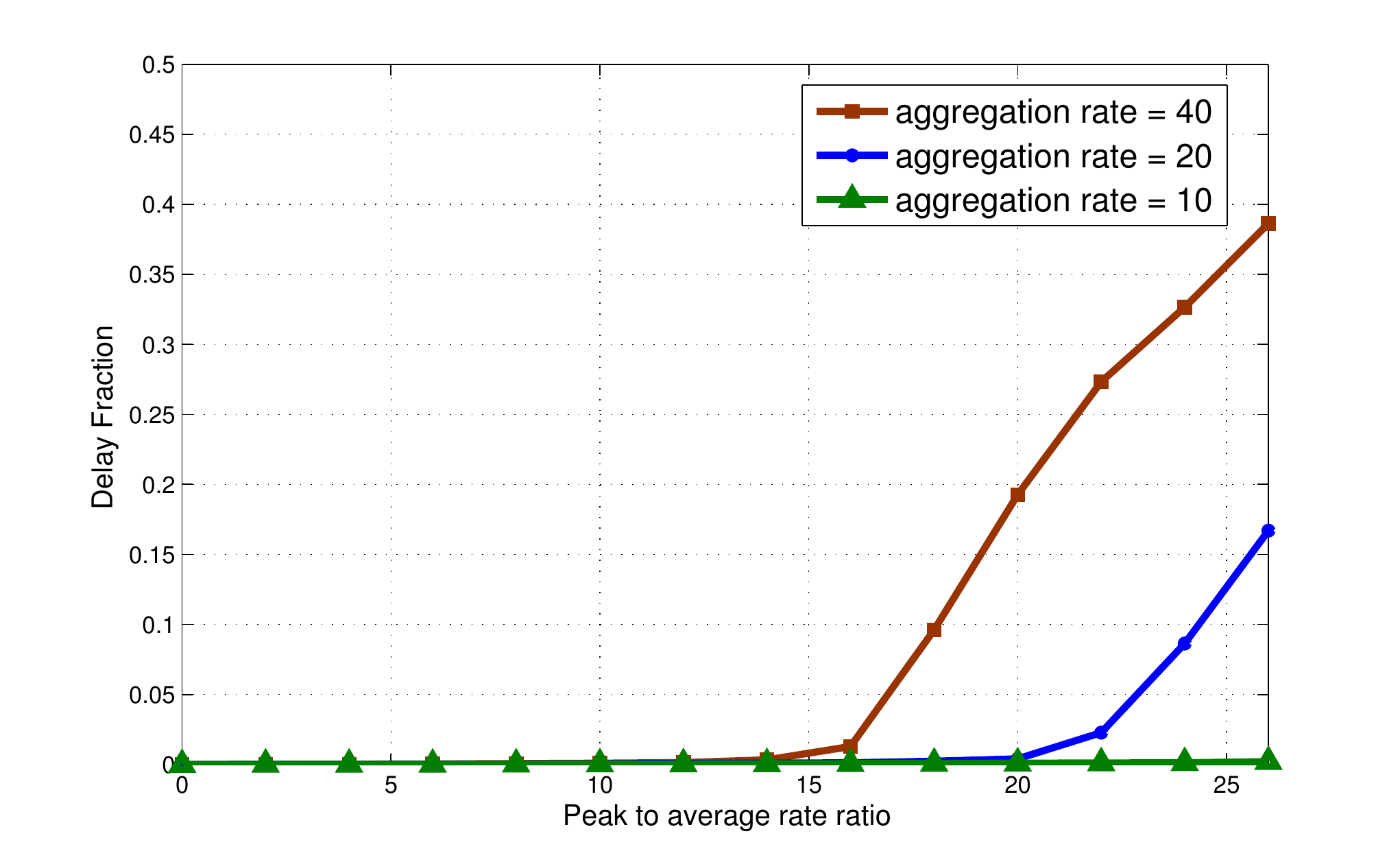}
                \caption{Analysis of performance under traffic burstiness}
                \label{fig7.2}
        \end{subfigure}
        \caption{Effect of traffic burstiness on MU-MIMO system performance. Zero peak to average rate ratio implies fully backlogged case}\label{fig7}
\end{figure*}

One of the important considerations for traffic burstiness, is to maintain high offered load. This ensures that the effect seen on the performance is due to timing alignment issue (discussed in detail in previous subsection) and not a lower offered load. For the case wherein the number of users is 12 and the AP serves 4 users at a time, the effect of traffic burstiness has been shown in fig.~\ref{fig7.1}. At aggregation rate of 10, the curve remains flat for all values of peak to average rate ratios which proves that the offered load provided to the system is high and the effect on the system performance is due to timing alignment and aggregation delay. The AP needs to serve 4 users at a time on a first in first out basis. At lower peak to average rate ratios, the AP is able to do this without a lot of delay and hence the throughput remains steady at lower values of peak to average rate ratios. As more and more frames are aggregated to amortize the overhead, the system has to bear the cost of the idle times between packet bursts to gather sufficient number of frames for atleast 4 users. This also increases the aggregation delay as a result of which throughput begins to decrease. Hence the throughput for aggregation rate of 40, falls at a lower peak to average rate ratio as compared to the throughput for aggregation rate 20. At aggregation rate of 40 the drop in throughput is as much as 40 Mbps at peak to average rate ratio of 27.

\subsubsection{Analysis of results}
The hypothesis is that the decline in performance is caused due to the delay the AP faces while waiting for packets for 4 users. So to formally evaluate the performance, we define a \emph{delay fraction} parameter defined as follows:
\\
\begin{equation}
Delay Fraction = \frac{\sum_{i=1}^{i=N} delay_{i}}{Time(total)}
\end{equation}
where  
\begin{equation}
delay_{i} = Tx_{current}(start) - Tx_{previous}(end) 
\end{equation}
\\
where $i$ denotes the $i^{th}$ transmit cycle and $Time(total)$ is the total transmission time (for the entire simulation). Thus, this parameter denotes the fraction of the total time the AP spent waiting for sufficient number of packets to transmit. If this fraction is greater than zero it would denote that the deterioration in performance is due to {\bf Case1} described previously. However, a value of zero implies {\bf Case2}. This is assuming that the offered load is sufficiently high. This parameter cannot directly distinguish between cases with higher offered loads and cases with lower offered loads because in both cases the AP would have to wait for packets. Therefore, a reference case would have to be taken to prove high offered load. 

For an aggregation rate of 10, as shown in \ref{fig7.2}, the delay fraction remains fixed at zero for all values of peak to average rate ratios. This would never happen if the offered load were below the critical value of offered load. Also since the timing delay problem would decrease with decreasing aggregation rate, the flat nature of the curve for aggregation rate of 10 proves that the offered load into the system is above the critical value for all peak to average rate ratios in the figure. Therefore, this serves as the reference case mentioned previously.

In this analysis, we answer the following questions:
\begin{itemize}
\item {\bf Why does the curve remain flat initially and what do the baseline values represent?} As shown in fig.~\ref{fig7.2}, the delay fraction remains fixed at zero in the region where the throughput curve is flat. This indicates that the flat region is an example of {\bf Case2} described previously. As the AP has packets for 4 users each time it is ready to transmit, this also represents a backlogged case. Also, since packet size variation and traffic burstiness were studied individually, no packet size variation has been considered. Hence, the base line values are essentially the throughput values for fixed packet scenario shown in fig.~\ref{fig6}. 
\item {\bf Why are the transition point where they are?} As the offered load has been maintained high enough for all values of peak to average rate ratios, the throughput essentially degrades due to the timing alignment. As a result, the transition happens when the scenario switches from {\bf Case2} to {\bf Case1}. This is indicated by the transition points for delay fraction in fig.~\ref{fig7.2}. As {\bf Case1} would be more prominent at higher aggregation rates, the throughput curve falls early for higher aggregation rates as compared to lower aggregation rates. 
\item {\bf Why is throughput independent of aggregation rates at higher peak to average rate ratios?} As shown in fig.~\ref{fig7.2}, the delay fraction for aggregation rate of 40 is extremely high as compared to aggregation rate of 20. As the throughput deterioration is more for higher values of delay fraction, the curves nearly coincide at higher peak to average rate ratios.  
\end{itemize}

%% file: secs/background.tex
\section{Related Work}
\label{sec:background}

\begin{itemize}
\item {\bf MU-MIMO performance evaluation under traffic dynamics:} Several works on performance evaluation of multi-user transmission have appeared in literature recently. Works such as \cite{nayak2017multi, nayak2019modeling, nayak2016performance, peshal2019modeling, nayakmodeling} model and evaluate the performance of MU-MIMO under the impact of closed loop dynamics. This is orthogonal to our focus which is on open loop traffic dynamics.  
\item {\bf Theoretical evaluation of MU-MIMO:} Excellent theoretical works such as \cite{3} and \cite{caire2010multiuser} highlight the performance benefits of MU-MIMO in terms of high throughput obtained from simultaneous data transmission. Although such works are necessary for a greater understanding, it is not possible for them to consider other factors that impact the performance. This gap is filled in by the research presented in this report.
\item {\bf Modifications and Improvements in existing protocol:} The papers in this category generally propose a new protocol/method by modifying on an existing one. For instance, in \cite{thapa2012mac} a protocol that switches between MU-MIMO and SU-MIMO so as to make the system resilient to link quality has been proposed whereas in \cite{bellalta2012performance} an improved aggregation technique to enhance the throughput of MU-MIMO transmissions has been proposed. However, to the best of my knowledge, no work related to performance improvement has ever considered the effect of variable traffic on MU-MIMO performance. This report highlights the need to improve MAC decisions in such protocols by accounting for traffic dynamics.
\item {\bf Hardware design studies} Works such as \cite{nayak2013multiband, nayak2014novel, nayak2013compact, nayak2012ultrawideband, nayak2021study, endluri975low, nayak2021novel, nayak2020novel, nayak2020design, nayak2020multiband, nayak2013multiband1, nayak2021triple, nayak2019asic, nayak2014underwater} have studied the impact of enhancing hardware design via smart and compact antenna designs. While an efficient antenna design is crucial to achieve a good system performance, the issues we highlight in this report need to be addressed via improvements to protocol design. 
\end{itemize}

%% file: secs/conclusion.tex
\section{Conclusion and Future Work}
In this report the performance evaluation of MU-MIMO under variable traffic has been presented. Packet size variations and traffic burstiness, the characteristic features of variable traffic, have been considered. Review of prior work indicates that such effects have not been accounted for previously and so this is the first work to account for variable traffic and its impact. The simulations indicate that the throughput gains promised by multi-user transmission are severely mitigated by traffic dynamics in certain scenarios even in absence of channel variation and mobility which would further deteriorate the performance. The fundamental reason for deterioration of system performance is due to simultaneous data transmission to multiple users. Therefore, a better approach for the AP would be to switch to a different transmit scheme such as MISO or to transmit to lesser number of users. Future work would focus on designing a MAC protocol robust to variable traffic. However, the key challenge lies in the fact that traffic characteristics are influenced by the interaction between the server and the user end. Hence, identifying a system lever parameter that would help identify the switching point is the key challenge.

%% file: main.bbl
\begin{thebibliography}{10}

\bibitem{2}
Robert Stacey.
\newblock Proposed specification framework for tgac.
\newblock {\em IEEE802. 11-09/9992r19}, 2011.

\bibitem{1}
Wolfgang John and Sven Tafvelin.
\newblock {Analysis of internet backbone traffic and header anomalies
  observed}.
\newblock In {\em Proceedings of the 7th ACM SIGCOMM conference on Internet
  measurement}, pages 111--116, 2007.

\bibitem{nayak2021ap}
Peshal Nayak.
\newblock {Ap-side WLAN Analytics}.
\newblock {\em arXiv preprint arXiv:2105.04524}, 2021.

\bibitem{nayak2021uscope}
Peshal Nayak and Edward~W Knightly.
\newblock {uScope: a Tool for Network Managers to Validate Delay-Based SLAs}.
\newblock In {\em Proceedings of the Twenty-second International Symposium on
  Theory, Algorithmic Foundations, and Protocol Design for Mobile Networks and
  Mobile Computing}, pages 171--180, 2021.

\bibitem{nayak2019virtual}
Peshal Nayak, Santosh Pandey, and Edward~W Knightly.
\newblock {Virtual Speed Test: an AP Tool for Passive Analysis of Wireless
  LANs}.
\newblock In {\em IEEE INFOCOM 2019-IEEE Conference on Computer
  Communications}, pages 2305--2313. IEEE, 2019.

\bibitem{nayakpassive}
Peshal Nayak.
\newblock {Passive AP-side Tool for WLAN Analysis: CAP Poster Abstract}.

\bibitem{nayak2017multi}
Peshal Nayak, Michele Garetto, and Edward~W Knightly.
\newblock {Multi-user downlink with single-user uplink can starve TCP}.
\newblock In {\em IEEE INFOCOM 2017-IEEE Conference on Computer
  Communications}, pages 1--9. IEEE, 2017.

\bibitem{nayak2019modeling}
Peshal Nayak, Michele Garetto, and Edward~W Knightly.
\newblock {Modeling Multi-User WLANs Under Closed-Loop Traffic}.
\newblock {\em IEEE/ACM Transactions on Networking}, 27(2):763--776, 2019.

\bibitem{nayak2016performance}
Peshal Nayak.
\newblock {\em {Performance Evaluation of MU-MIMO WLANs Under the Impact of
  Traffic Dynamics}}.
\newblock PhD thesis, 2016.

\bibitem{peshal2019modeling}
Nayak Peshal, Michele Garetto, Knightly Edward, et~al.
\newblock {Modeling Multi-User WLANs Under Closed-Loop Traffic}.
\newblock 2019.

\bibitem{nayakmodeling}
Peshal Nayak.
\newblock {Modeling and Performance Evaluation of MU-MIMO Under the Impact of
  Closed Loop Traffic Dynamics: CAP Poster Abstract}.

\bibitem{3}
Taesang Yoo and Andrea Goldsmith.
\newblock On the optimality of multiantenna broadcast scheduling using
  zero-forcing beamforming.
\newblock {\em IEEE Journal on selected areas in communications},
  24(3):528--541, 2006.

\bibitem{caire2010multiuser}
Giuseppe Caire, Nihar Jindal, Mari Kobayashi, and Niranjay Ravindran.
\newblock Multiuser mimo achievable rates with downlink training and channel
  state feedback.
\newblock {\em IEEE Transactions on Information Theory}, 56(6):2845--2866,
  2010.

\bibitem{thapa2012mac}
Anup Thapa and Seokjoo Shin.
\newblock A mac protocol to select optimal transmission mode in very high
  throughput wlan: Mu-mimo vs. multiple su-mimo.
\newblock In {\em 2012 Third Asian Himalayas International Conference on
  Internet}, pages 1--5. IEEE, 2012.

\bibitem{bellalta2012performance}
Boris Bellalta, Jaume Barcelo, Dirk Staehle, Alexey Vinel, and Miquel Oliver.
\newblock On the performance of packet aggregation in ieee 802.11 ac mu-mimo
  wlans.
\newblock {\em IEEE Communications Letters}, 16(10):1588--1591, 2012.

\bibitem{nayak2013multiband}
Peshal~B Nayak, Sudhanshu Verma, and Preetam Kumar.
\newblock {Multiband fractal antenna design for Cognitive radio applications}.
\newblock In {\em 2013 International Conference on Signal Processing and
  Communication (ICSC)}, pages 115--120. IEEE, 2013.

\bibitem{nayak2014novel}
Peshal~B Nayak, Sudhanshu Verma, and Preetam Kumar.
\newblock {A novel compact tri-band antenna design for WiMax, WLAN and
  bluetooth applications}.
\newblock In {\em 2014 Twentieth National Conference on Communications (NCC)},
  pages 1--6. IEEE, 2014.

\bibitem{nayak2013compact}
Peshal~B Nayak, Ramu Endluri, Sudhanshu Verma, and Preetam Kumar.
\newblock {Compact dual-band antenna for WLAN applications}.
\newblock In {\em 2013 IEEE 24th Annual International Symposium on Personal,
  Indoor, and Mobile Radio Communications (PIMRC)}, pages 1381--1385. IEEE,
  2013.

\bibitem{nayak2012ultrawideband}
Peshal~B Nayak, Sudhanshu Verma, and Preetam Kumar.
\newblock {Ultrawideband (UWB) antenna design for cognitive radio}.
\newblock In {\em 2012 5th International Conference on Computers and Devices
  for Communication (CODEC)}, pages 1--4. IEEE, 2012.

\bibitem{nayak2021study}
Peshal Nayak, Sudhanshu Verma, and Preetam Kumar.
\newblock {A Study of Ultrawideband (UWB) Antenna Design for Cognitive Radio
  Applications}.
\newblock {\em arXiv preprint arXiv:2106.15272}, 2021.

\bibitem{endluri975low}
Ramu Endluri, Peshal~B Nayak, and Preetam Kumar.
\newblock {A Low Cost Dual Band Antenna for Bluetooth, 2.3 GHz WiMAX and
  2.4/5.2/5.8 GHz WLAN}.
\newblock {\em International Journal of Computer Applications}, 975:8887.

\bibitem{nayak2021novel}
Peshal~B Nayak, Ramu Endluri, Sudhanshu Verma, and Preetam Kumar.
\newblock {A Novel Compact Dual-Band Antenna Design for WLAN Applications}.
\newblock {\em arXiv preprint arXiv:2106.13232}, 2021.

\bibitem{nayak2020novel}
Peshal~B Nayak.
\newblock {A Novel Triple Band Antenna Design for Bluetooth, WLAN and WiMAX
  Applications}.
\newblock 2020.

\bibitem{nayak2020design}
Peshal~B Nayak.
\newblock {Design and Analysis of a Multiband Fractal Antenna for Applications
  in Cognitive Radio Technologies}.
\newblock 2020.

\bibitem{nayak2020multiband}
Peshal~B Nayak.
\newblock {Multiband Fractal Antenna Designs}.
\newblock 2020.

\bibitem{nayak2013multiband1}
Peshal Nayak.
\newblock {Multiband Antenna Designs for Wireless Communication Systems}.
\newblock 2013.

\bibitem{nayak2021triple}
Peshal Nayak.
\newblock {Triple Band Antenna Design for Bluetooth, WLAN and WiMAX
  Applications}.
\newblock 2021.

\bibitem{nayak2019asic}
Peshal Nayak.
\newblock {ASIC Technology Roadmap: A Literature Review}.
\newblock {\em Health}, 2(3):1, 2019.

\bibitem{nayak2014underwater}
Peshal Nayak.
\newblock {\em {Underwater Communications for UUV System}}.
\newblock PhD thesis, Indian Institute of Technology Patna, 2014.

\end{thebibliography}
